\documentclass[preprintnumbers, floatfix, preprintnumbers, letterpaper, superscriptaddress,nofootinbib]{revtex4}
\pdfoutput=1
\usepackage{graphicx}
\usepackage{microtype}
\usepackage{amsmath}
\usepackage{amssymb}
\usepackage{subfigure}
\usepackage{hyperref}
\usepackage{url}
\usepackage{xcolor}
\usepackage{color}
\usepackage{mathrsfs}
\usepackage{calrsfs}
\usepackage{amsfonts}
\usepackage{latexsym}
\usepackage{ragged2e}
\usepackage{epsfig}
\usepackage{textcomp}
\usepackage{phaistos}
\usepackage[utf8]{inputenc}
\makeatletter
\renewcommand\@makefnmark{\hbox{\@textsuperscript{\normalfont\color{purple}\@thefnmark}}}
\renewcommand\@makefntext[1]{%
  \parindent 1em\noindent
            \hb@xt@1.8em{%
                \hss\@textsuperscript{\normalfont\@thefnmark}}#1}
\makeatother

\usepackage{caption}
\DeclareCaptionJustification{justified}{\leftskip=0pt \rightskip=0pt \parfillskip=0pt plus 1fil}
\definecolor{vividviolet}{rgb}{0.62, 0.0, 1.0}
\definecolor{amaranth}{rgb}{0.9, 0.17, 0.31}
\definecolor{palatinateblue}{rgb}{0.15, 0.23, 0.89}
\definecolor{brightpink}{rgb}{1.0, 0.0, 0.5}
\definecolor{cornflowerblue}{rgb}{0.39, 0.58, 0.93}
\definecolor{deepcarminepink}{rgb}{0.94, 0.19, 0.22}
\definecolor{radicalred}{rgb}{1.0, 0.21, 0.37}

\hypersetup{ linktoc=all,
    colorlinks, linkcolor={palatinateblue},
    citecolor={brightpink}, urlcolor={amaranth}
}

\graphicspath{{Images/}}

\def\sideremark#1{\ifvmode\leavevmode\fi\vadjust{\vbox to0pt{\vss
 \hbox to 0pt{\hskip\hsize\hskip1em
 \vbox{\hsize1.5cm\tiny\raggedright\pretolerance10000
 \noindent #1\hfill}\hss}\vbox to8pt{\vfil}\vss}}}%
\begin{document}

\title{Ellis Wormholes in Anti-De Sitter Space}

\author{Jose Luis \surname{Bl\'azquez-Salcedo}}
\email{jlblaz01@ucm.es}
\affiliation{Departamento de F\'isica Te\'orica, Universidad Complutense de Madrid, E-28040 Madrid, Spain}

\author{Xiao Yan \surname{Chew}}
\email{xychew998@gmail.com}
\affiliation{Department of Physics Education, Pusan National University, Busan 46241, Republic of Korea}
\affiliation{Research Center for Dielectric and Advanced Matter Physics, Pusan National University, Busan 46241, Republic of Korea}

\author{Jutta \surname{Kunz}}
\email{jutta.kunz@uni-oldenburg.de}
\affiliation{Institut f\"ur  Physik, Universit\"at Oldenburg, Postfach 2503, D-26111 Oldenburg, Germany}

\author{Dong-han \surname{Yeom}}
\email{innocent.yeom@gmail.com}
\affiliation{Department of Physics Education, Pusan National University, Busan 46241, Republic of Korea}
\affiliation{Research Center for Dielectric and Advanced Matter Physics, Pusan National University, Busan 46241, Republic of Korea}

\begin{abstract}
We construct traversable wormholes with anti-de Sitter asymptotics supported by a phantom field. These wormholes are massless and symmetric with respect to reflection of the radial coordinate $\eta \to - \eta$. Their circumferential radius decreases monotonically from radial infinity to their single throat. Analogous to their asymptotically flat counterparts, these anti-de Sitter wormholes possess an unstable radial mode.
\end{abstract}

\maketitle

\section{Introduction}

In order to resolve the information loss paradox of black holes \cite{Hawking:1976ra}, there have been numerous candidate proposals (see e.g. \cite{Chen:2014jwq} for a review), but many of them were not successful and several candidates have been critized, eg., black hole conplementarity \cite{Yeom:2008qw, Yeom:2009zp} or the firewall conjecture \cite{Chen:2015gux}. In this regard, one of the possible alternatives is to introduce non-local interactions between the inside and outside of the horizon \cite{Yeom:2016qec,Hwang:2017yxp}. In order to realize this idea without violating known principles of quantum mechanics, the so-called the ER=EPR conjecture has been proposed \cite{Maldacena:2013xja}, where this idea is partly supported by holography \cite{Almheiri:2019hni}. According to the conjecture, for a given entangled particle (Einstein-Podolsky-Rosen; EPR), there exists its dual geometry as the Einstein-Rosen (ER) bridge, which is the most well-known example of non--traversable wormholes in general relativity (GR), where it can be naturally obtained from the Schwarzschild black hole by maximally extending the solution \cite{Einstein:1935tc}. This implies that through the ER bridge, there can exist a conspiracy between the earlier part of Hawking radiation and the inside degrees of freedom.

When the ER=EPR conjecture was first proposed, people assumed that the ER bridge must be non-traversable, because there is no information transfer via EPR pairs. However, later it was noticed that the ER bridge can become traversable if one considers non-perturbative effects \cite{Chen:2016nvj} or correlations between two asymptotic boundaries \cite{Maldacena:2018lmt}. If one constructs a traversable wormhole from a well-defined dual field theory, it might be consistent with the ER=EPR conjecture. On the other hand, if a traversable wormhole can be obtained from a more classical or semi-classical construction, it may indicate that the original proposal of the ER=EPR conjecture may be inconsistent or the proposal should be restricted. In order to do this, the violation of the averaged null energy condition would be required \cite{Hartman:2016lgu}, which would be possible if one did not consider the expectation value of the energy-momentum tensor of the entire histories \cite{Chen:2016nvj} (i.e., if one would consider only a specific non-perturbative process \cite{Chen:2015gux,Sasaki:2014spa,Chen:2017suz,Chen:2018aij}) or if the theory would violate causality a little due to modifications of gravity (e.g., 
\cite{Hochberg:1990is,Fukutaka:1989zb,Ghoroku:1992tz,Furey:2004rq,Bronnikov:2009az, Kanti:2011jz,Sushkov:2011jh,Kanti:2011yv,Antoniou:2019awm,Ibadov:2020btp}).

Keeping these motivations in mind, we note that there are various models of wormholes that connect asymptotically locally AdS spaces \cite{Lemos:2003jb,Korolev:2014hwa,Anabalon:2018rzq,Franciolini:2018aad,Mironov:2018uou,Anabalon:2020loe,Nozawa:2020gzz,Chatzifotis:2020oqr}. However in this work, we focus at obtaining wormhole with AdS asymptotics, which correspond to the AdS generalization of the asymptotically flat Ellis wormholes \cite{Ellis:1973yv,Ellis:1979bh,Bronnikov:1973fh}. These represent analytically constructed wormhole solutions in GR, where the throat is supported by a phantom field, which is a real--valued scalar field that has an opposite sign of the kinetic term. Thus, it is a simple type of exotic matter that violates the null energy condition. This is necessary for the construction of traversable wormholes in GR, in order to prevent the throat from collapse (see e.g.~\cite{Visser:1995cc}). 
This type of exotic matter can be used to explain the accelerated expansion of the universe \cite{Caldwell:1999ew,Carroll:2003st,Gibbons:2003yj,Hannestad:2005fg}.  Besides the Ellis wormholes, which might be the simplest example for traversable and asymptotically flat wormholes in GR \cite{Ellis:1973yv,Ellis:1979bh,Bronnikov:1973fh}, a phantom field has also be employed to construct black holes \cite{Bronnikov:2005gm,Chen:2016yey}, 4--dimensional black rings \cite{Kleihaus:2019wck}, or star--like objects \cite{Dzhunushaliev:2008bq}.

The Ellis wormhole has a very simple geometrical structure, it possesses only a single throat that connects two asymptotically flat regions. It violates the averaged null energy condition, and possesses an unstable radial mode \cite{Gonzalez:2008wd,Gonzalez:2008xk,Torii:2013xba,Bronnikov:2012ch}. It has been generalized to higher dimensions \cite{Torii:2013xba}, and to the rotating case in four dimensions \cite{Kashargin:2007mm,Kashargin:2008pk,Kleihaus:2014dla,Chew:2016epf} and five dimensions \cite{Dzhunushaliev:2013jja}. In addition, properties of rotating Ellis wormholes have been studied in modified gravity, e.g., in scalar--tensor theory \cite{Chew:2018vjp}. It has also been shown that a phantom field may no longer be necessary for the construction of wormholes, when the gravity sector is suitably modified \cite{Hochberg:1990is,Fukutaka:1989zb,Ghoroku:1992tz,Furey:2004rq,Bronnikov:2009az, Kanti:2011jz,Sushkov:2011jh,Kanti:2011yv,Antoniou:2019awm,Ibadov:2020btp}. 
{
In fact, wormholes in GR can also be supported only by fermions \cite{Blazquez-Salcedo:2019uqq}, and recently, traversable wormholes have been constructed in the Einstein--Maxwell--Dirac system \cite{Blazquez-Salcedo:2020czn}.
}

In addition, configurations of wormholes with mixed phantom and ordinary fields have been constructed, for example, Ellis wormholes immersed in bosonic matter \cite{Dzhunushaliev:2014bya,Hoffmann:2017jfs,Hoffmann:2017vkf,Hoffmann:2018oml} or mixed neutron star--wormhole systems \cite{Dzhunushaliev:2011xx,Dzhunushaliev:2012ke,Dzhunushaliev:2013lna,Dzhunushaliev:2014mza,Aringazin:2014rva}. Since wormholes can represent compact objects that might mimic black holes, in order to distinguish them from black holes, a number of astrophysical signatures of wormholes have been pointed out that might allow to search for their existence in the near future. Examples are their shadows \cite{Nedkova:2013msa,Gyulchev:2018fmd,Amir:2018szm}, gravitational lensing \cite{Abe:2010ap,Toki:2011zu,Takahashi:2013jqa,Cramer:1994qj,Perlick:2003vg,Tsukamoto:2012xs,Bambi:2013nla}, accretion disks around wormholes \cite{Zhou:2016koy}, and their ringdown phase with the associated emission of gravitational waves \cite{Blazquez-Salcedo:2018ipc}.

This paper is organized as follows. In Sec.~\ref{sec:th}, we briefly introduce our theoretical setup comprising the phantom field and the metric ansatz. In addition, we derive the set of coupled differential equations, we describe the numerical methods employed to solve these equations, and we study the asymptotic behavior of the metric functions. In particular, we discuss the mass of the wormholes in asymptotically AdS spacetimes and the geometric properties of the wormholes, as well as the violation of the energy conditions. In Sec.~\ref{sec:res}, we present and discuss our numerical results for the wormhole  solutions. In Sec.~\ref{sec:lin}, we study the stability of the wormholes by calculating the unstable mode of the radial perturbations of the metric and the phantom field. Finally, in Sec.~\ref{sec:con}, we summarize our work and present an outlook.

\section{\label{sec:th}Theoretical Setting}

\subsection{Theory}

We consider the Einstein--Hilbert action including the cosmological constant $\Lambda$ and the Lagrangian for the matter field $\mathcal{L}_{m}$ 
\begin{equation} \label{EHaction}
 S_{\text{EH}}=  \int d^4 x \sqrt{-g}  \left[  \frac{1}{16 \pi G} (R-2\Lambda) + \mathcal{L}_{\text{ph}} \right]  \,,
\end{equation}
where $\Lambda$ is related to the AdS length $l$ by $\Lambda=\pm 3/l^2$, and $ \mathcal{L}_{\text{ph}}$ is the Lagrangian of the phantom field $\psi$,
\begin{equation}
  \mathcal{L}_{\text{ph}} = \frac{1}{2} \partial_\mu \psi \partial^\mu \psi \,.
\end{equation} 

By varying the action with respect to the metric, we obtain the Einstein equations,
\begin{equation} \label{einstein_eqn}
 R_{\mu \nu} - \frac{1}{2} g_{\mu \nu} R + \Lambda g_{\mu \nu} = 2  \kappa T_{\mu \nu} \,,
\end{equation}
where $\kappa=4 \pi G$, and the stress--energy tensor $T_{\mu \nu}$ is given by
\begin{equation}
 T_{\mu \nu} = \frac{1}{2} g_{\mu \nu} \partial_\alpha \psi \partial^\alpha \psi -  \partial_\mu \psi \partial_\nu \psi \,.
\end{equation}
%
We obtain the massless Klein--Gordon equation by varying with respect to the phantom field,
\begin{equation} \label{KGeqn}
  \frac{1}{\sqrt{-g}}  \partial_\mu \left(  \sqrt{-g} \partial^\mu \psi  \right)  = 0 \,.
\end{equation}

In order to construct wormhole solutions with AdS asymptotics, we employ the following line element with a quasi--isotropic radial coordinate $\eta$,
\begin{equation}  \label{line_element}
 ds^2 = -F(\eta) N(\eta) dt^2 + \frac{p(\eta)}{F(\eta) } \left[ \frac{d\eta^2}{N(\eta)} + h(\eta) (d \theta^2+\sin^2 \theta d\varphi^2)   \right]\,,
\end{equation}
where $N(\eta)= 1 -  \Lambda \eta^2/3 $ and $h(\eta)=\eta^2+\eta_0^2$ with $\eta_0$ the throat parameter. In pure Einstein gravity $(\Lambda=0)$, the above metric describes a static Ellis wormhole, that possesses two asymptotically flat regions $\Sigma_{\pm}$ as $\eta \rightarrow \pm \infty$. The analytical solution for static Ellis wormholes is given by
\begin{equation}
 p(\eta) = 1 \,, \quad F(\eta) = e^{f(\eta)} \,, 
\end{equation}
with
\begin{equation}
 f(\eta) = \frac{2 M}{\eta_0} \left[  \arctan \left( \frac{\eta}{\eta_0} \right)-\frac{\pi}{2}    \right] \,,
\end{equation}
and $M$ is the mass of the Ellis wormholes on $\Sigma_{+}$. However, when $\Lambda < 0$
the above metric should possess two asymptotically AdS regions as $\eta \rightarrow \pm \infty$.

\subsection{Ordinary Differential Equations (ODEs)}

By substituting the line element Eq.~\eqref{line_element} into the Einstein equations, we obtain a set of second--order nonlinear ODEs for the metric functions,
\begin{align}
F'' - \frac{F}{p} p'' + \left[ \frac{p'}{2 p} + \frac{\eta (6 N- h \Lambda)}{3 h N} \right] F' - \frac{ \eta F (6 N-h \Lambda)}{3 p h N} p' + \frac{ 3 F}{ 4 p^2} p'^2  - \frac{5}{4 F} F'^2 & \nonumber \\
- \frac{6 h F N+3 \Lambda p h^2-2 h \Lambda \eta^2 F-3 h F-3 \eta^2 F N}{3 N h^2}  &= - \kappa F \psi'^2 \,,  \label{ode1} \\
\frac{\Lambda \eta}{3 F N} F' + \frac{\eta (3 N-h \Lambda)}{3 p h N} p' + \frac{p'^2}{4 p^2} - \frac{F'^2}{4 F^2} +   \frac{3 N \eta^2 F +3 \Lambda p h^2-3 h F-2 h \Lambda \eta^2 F}{3 F h^2 N} &= - \kappa \psi'^2 \,, \label{ode2} \\
p'' + \frac{\eta (3 N-h \Lambda)}{3 h N} p' - \frac{2 \Lambda \eta p}{3 F N} F' - \frac{p'^2}{p} + \frac{p}{2 F^2} F'^2 \qquad \qquad \qquad \qquad \qquad  \qquad & \nonumber \\
+ \frac{2 p (3 h F N-2 h \Lambda \eta^2 F+3 \Lambda p h^2-F h^2 \Lambda-3 \eta^2 F N)}{3 F N h^2} &= 2 \kappa p \psi'^2  \label{ode3} \,,
\end{align}
where the prime denotes the derivative of the functions with respect to the radial coordinate $\eta$. 

From the massless Klein--Gordon equation Eq.~\eqref{KGeqn} for the phantom field we obtain a first integral, 
\begin{equation} \label{odepsi}
 \psi' =  \frac{D}{ h N \sqrt{p}} \,,
\end{equation}
where $D$ is a constant. In the asymptotically flat case it is interpreted as the scalar charge of the phantom field. 
We eliminate the term $\psi'^2$ in Eqs.~\eqref{ode1} and \eqref{ode3} by adding Eq.~\eqref{ode1} and \eqref{ode3}, respectively, to Eq.~\eqref{ode2}. Then we obtain the following ODEs for the metric functions $f$ and $p$,
\begin{align}
 F'' &= \frac{F'^2}{F} + \frac{\Lambda \eta F}{3 p N} p' - \left( \frac{p'}{2 p}+\frac{2 \eta (3 N-h \Lambda)}{3 h N} \right) F'  +  \frac{2 \Lambda (2 \eta^2 F+F h-3 p h)}{3 N h} \,,   \label{odeF} \\
p'' &= \frac{p'^2}{2 p} - \frac{ \eta (3 N-h \Lambda)}{N h} p' +\frac{2 p (3 F - 3 F N + 4 \Lambda \eta^2 F + \Lambda h F- 6 \Lambda p h)}{3 h F N} \,.  \label{odep}
\end{align} 

In order to study the asymptotic behavior of the metric functions in the limit $\eta \rightarrow \infty$, we perform the series expansion for Eqs.~\eqref{odeF} and \eqref{odep} to obtain the asymptotic expansion for the functions
 \begin{align} \label{large_eta_expan1}
 F(\eta) &= F_\infty + \frac{F_\infty \eta_0^2}{3 \eta^2} - \frac{F_\infty \eta^2_0 \left( \Lambda \eta_0^2-12  \right)}{15 \Lambda \eta^4}       + O (\eta^{-6})  \,, \\
 p(\eta) &= F_\infty - \frac{F_\infty \eta_0^2}{3 \eta^2}  +  \frac{F_\infty \eta^2_0 \left( 14 \Lambda \eta_0^2 + 27  \right)}{45 \Lambda \eta^4}        + O (\eta^{-6})  \,.
\end{align}
We observe that the odd terms vanish identically. With these expansions, the large-$\eta$ expansions of $g_{tt}$  and $g^{\eta \eta}$ for the wormhole are given by
\begin{align}
- g_{tt} \Big|_{\eta \rightarrow \infty} &=  -\frac{\Lambda F_\infty \eta^2}{3} + F_\infty \left( 1-\frac{\Lambda \eta^2_0}{9}  \right)  + \frac{F_\infty \eta^2_0}{15 \eta^2} \left( 1+ \frac{\Lambda \eta^2_0}{3}    \right) + O (\eta^{-4})  \,, \\
 g^{\eta \eta}  \Big|_{\eta \rightarrow \infty} &= -\frac{\Lambda \eta^2}{3} + 1 - \frac{2 \Lambda \eta^2_0}{9} + \frac{\eta^2_0 \left( 7 \Lambda \eta^2_0 + 81  \right)}{135 \eta^2}  +  O (\eta^{-4})  \,.  \label{large_eta_expan2}
\end{align}
Since the odd terms vanish, we find that the metric functions at $\eta \rightarrow -\infty$ have exactly the same asymptotic expansions as the metric functions at $\eta \rightarrow \infty$. 
According to the asymptotic expansions, the appropriate boundary conditions to be imposed on the metric functions at infinity are given by   
\begin{equation}
 F (\pm \infty) = p (\pm \infty)  = 1\,,
\end{equation}  
and thus $F_\infty=1$.

Considering the expression for the mass of the wormholes as, for instance, 
obtained from the Ashtekar--Magnon--Das formalism \cite{Ashtekar:1984zz,Ashtekar:1999jx,Das:2000cu}
the vanishing of the odd terms also implies that the mass of these symmetric wormholes vanishes,
as it happens in the symmetric case also for an asymptotically flat space.

In additon, the series expansion at $\eta=0$ is given by
\begin{align}
 F(\eta) &= F_0 + \frac{\Lambda}{3} (F_0-3 p_0) \eta^2 +  O (\eta^4)    \,, \\
 p(\eta) &=  p_0 + \frac{\Lambda p_0 (F_0 - 6 p_0)}{3 F_0} \eta^2 + O (\eta^4) \,.
\end{align}

By using Eq.~\eqref{odepsi}, we may rewrite Eq.~\eqref{ode2} as 
\begin{equation}
D^2 = \frac{N^2 p h^2}{4 F^2} F'^2 -\frac{ N p \Lambda \eta h^2}{3 F} F' -  \frac{ N^2 h^2}{4 p} p'^2 - \frac{1}{3} N \eta h (3 N-h \Lambda) p' -\frac{ N p (3 N \eta^2 F+3 \Lambda p h^2-3 F h-2 h \Lambda \eta^2 F)}{3 F}\,.
\end{equation}
This equation can be used to monitor the accuracy of the numerical computation by ensuring that $D$ is constant on the grid in the full domain of integration. We note that in GR without a cosmological constant, for a generic static Ellis wormhole the mass $M$ and the scalar charge $D$ are related by
\begin{equation}
 D^2 = M^2 + \eta_0^2 \,.
\end{equation}

We solve Eqs.~\eqref{odeF} and \eqref{odep} numerically by using the ODE solver package COLSYS, which tackles boundary value problems for systems of nonlinear coupled ODEs based on the Newton--Raphson method \cite{Ascher:1979iha}. Employing an adaptive grid selection procedure and using more than 1000 points, COLSYS provides the solutions with high accuracy together with an error estimate. 
To integrate the ODEs in the full interval $-\infty$ to $+\infty$, we compactify the radial coordinate $\eta$ in the numerical calculations as follows: $\eta=\eta_0 \tan (\pi x/2)$ with $x \in [-1,1]$. Moreover, we change to dimensionless variables by introducing the following rescaled variables/parameters in the above ODEs, 
\begin{equation}
\tilde{\psi}= \sqrt{\kappa} \psi \,, \quad \eta = \eta_0 \tilde{\eta} \,, \quad \Lambda = \frac{\tilde{\Lambda}}{\eta_0^2}\,.
\label{scaling}
\end{equation}
In the remaining sections we will omit the tilde for convenience. After the rescaling, the only free parameter left is $\Lambda$.

\subsection{Geometric Properties}

To study the geometric properties of the wormholes, we now define 
the circumferential radius $R_c$ as
\begin{equation}
 R_c (\eta) = \sqrt{ \frac{p h}{F} } \,.
\end{equation}
A wormhole throat is a minimal surface of the wormhole. Therefore it has to satisfy the following conditions,
\begin{equation}
 \frac{d R_c}{d \eta} \Bigg|_{\eta=\eta_{\text{th}}} = 0\,, \qquad \frac{d^2 R_c}{d \eta^2}  \Bigg|_{\eta=\eta_{\text{th}}} > 0 \,,
\end{equation}
where $\eta_{\text{th}}$ is the radial coordinate of the throat.
If a wormhole has an equator with radial coordinate $\eta_{\text{eq}}$, this requires
\begin{equation}
 \frac{d R_c}{d \eta} \Bigg|_{\eta=\eta_{\text{eq}}} = 0\,, \qquad \frac{d^2 R_c}{d \eta^2} \Bigg|_{\eta=\eta_{\text{eq}}} < 0 \ .
\end{equation}
An equator is typically located between two throats, and thus signals a double throat configuration. 

A wormhole throat can be visualized 
by embedding the equatorial plane $(\theta=\pi/2)$ in Euclidean space $(\rho,\varphi,z)$.
Using cylindrical coordinates, this implies for the above metric parametrization
\begin{align}
ds^2 &= \frac{p}{F N}  d \eta^2 + \frac{p h}{F}   d \phi^2 \, \\
&= d \rho^2 + dz^2 + \rho^2 d \phi^2   \,.
\end{align}
We then obtain the expression for $z$ by comparison,
\begin{equation} \label{formula_embedding}
  z =  \pm \int  \sqrt{  \frac{p}{F N}  -   \left( \frac{d \rho}{d \eta} \right)^2    }     d \eta \,, \quad \rho \equiv R_c \,, 
\end{equation}
where the sign of $z$ depends on the sign of the radial coordinate $\eta$. For the special case of the massless symmetric Ellis wormhole without cosmological constant $(\Lambda=0,F=p=1)$ this yields
\begin{equation}
z =  \int_{0}^{\eta} \sqrt{  1  -  \frac{{\eta'}^2}{{\eta'}^2+{\eta'}_0^2}     }   \,  d \eta' = \text{arcsinh} \left( \frac{\eta}{\eta_0} \right) \,.
\end{equation}

\subsection{Null Energy Condition (NEC)}

As noted above, the construction of wormholes requires the violation of the energy conditions. Here, we focus on the NEC, which states that
\begin{equation}
  T_{\mu \nu} k^\mu k^\nu \geq 0 \,,
\end{equation}
for all (future--pointing) null vectors $k_\mu$ which satisfy $k_\mu k^\mu=0$. We note that the violation of the NEC also implies the violation of the weak and strong energy conditions.

Since the wormhole spacetime is spherically symmetric, there are two choices of null vector \cite{Antoniou:2019awm}, 
\begin{equation}
k_\mu = \left( g_{tt} , \sqrt{-\frac{g_{tt}}{g_{\eta \eta}}}, 0, 0 \right) \,, \quad \text{and} \quad k_\mu = \left( 1, 0, \sqrt{- \frac{g_{tt}}{g_{\theta \theta}}  } ,  0 \right)\,,
\end{equation}
which yield two expressions to test NEC violation,
\begin{equation}
 -T^t\,_t + T^\eta\,_\eta \geq 0 \,, \quad  -T^t\,_t + T^\theta\,_\theta \geq 0 \,.
\end{equation}
Evaluating the above expressions explicitly, 
\begin{equation}
-T^t\,_t + T^\eta\,_\eta = - \frac{D^2 F}{p^2 h^2 N}<0\,, \quad  -T^t\,_t + T^\theta\,_\theta = 0 \,,
\label{necvio}
\end{equation}
shows that the NEC is always violated.

\begin{figure}[t!]
\centering
\mbox{ 
(a)  \includegraphics[angle =-90,scale=0.3]{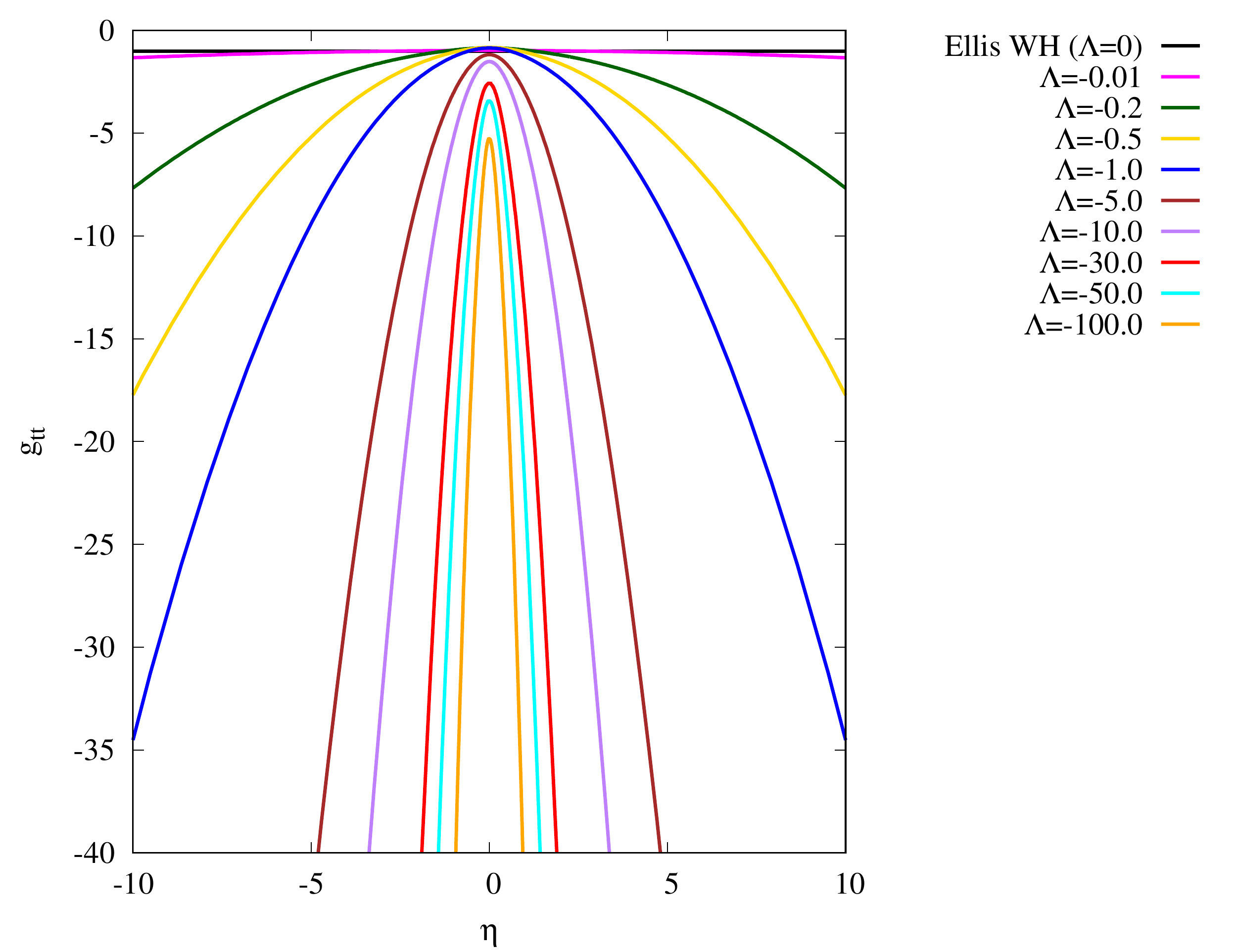}
(b) \includegraphics[angle =-90,scale=0.3]{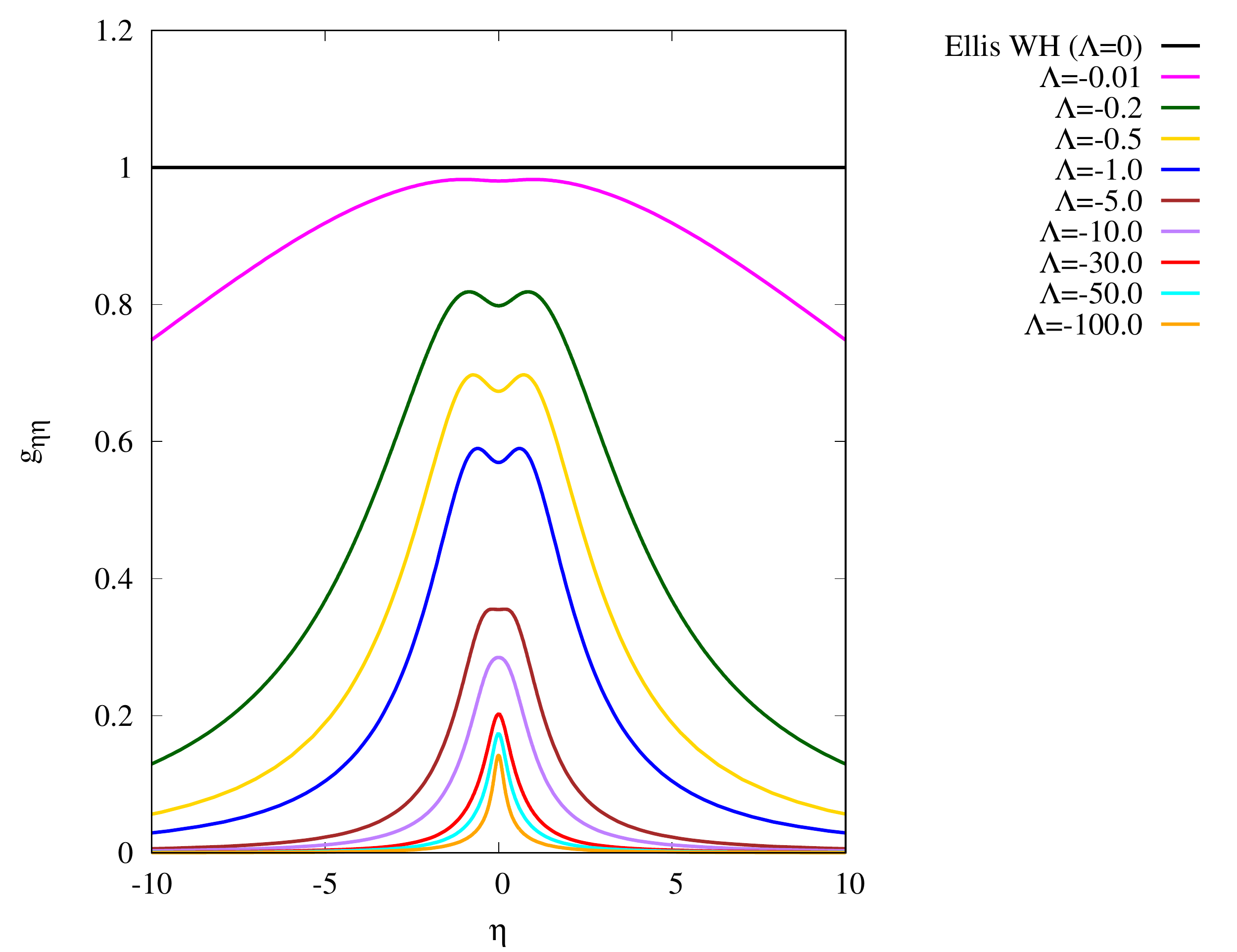} 
}
\mbox{
(c)   \includegraphics[angle =-90,scale=0.3]{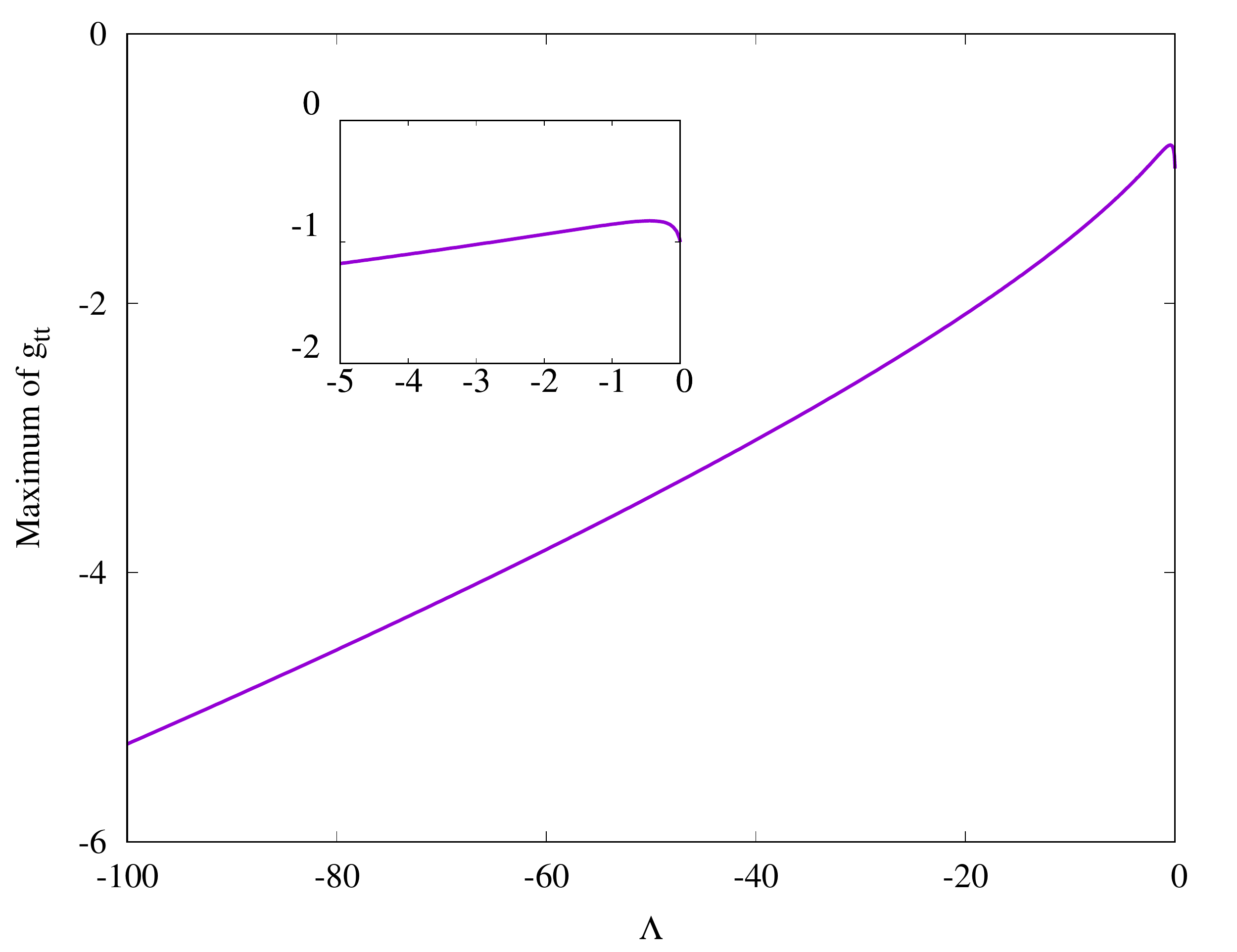}
}
\caption{Metric functions (a) $g_{tt}$ and (b) $g_{\eta \eta}$ vs the radial coordinate $\eta$ for $-100 \leq \Lambda \leq 0$; (c) global maximum of $g_{tt}$ vs the cosmological constant $\Lambda$.} 
\label{plot_metric}
\end{figure}

\section{\label{sec:res}Results and Discussion}

\begin{figure}[t!]
\centering
\mbox{ 
(a)  \includegraphics[angle =-90,scale=0.3]{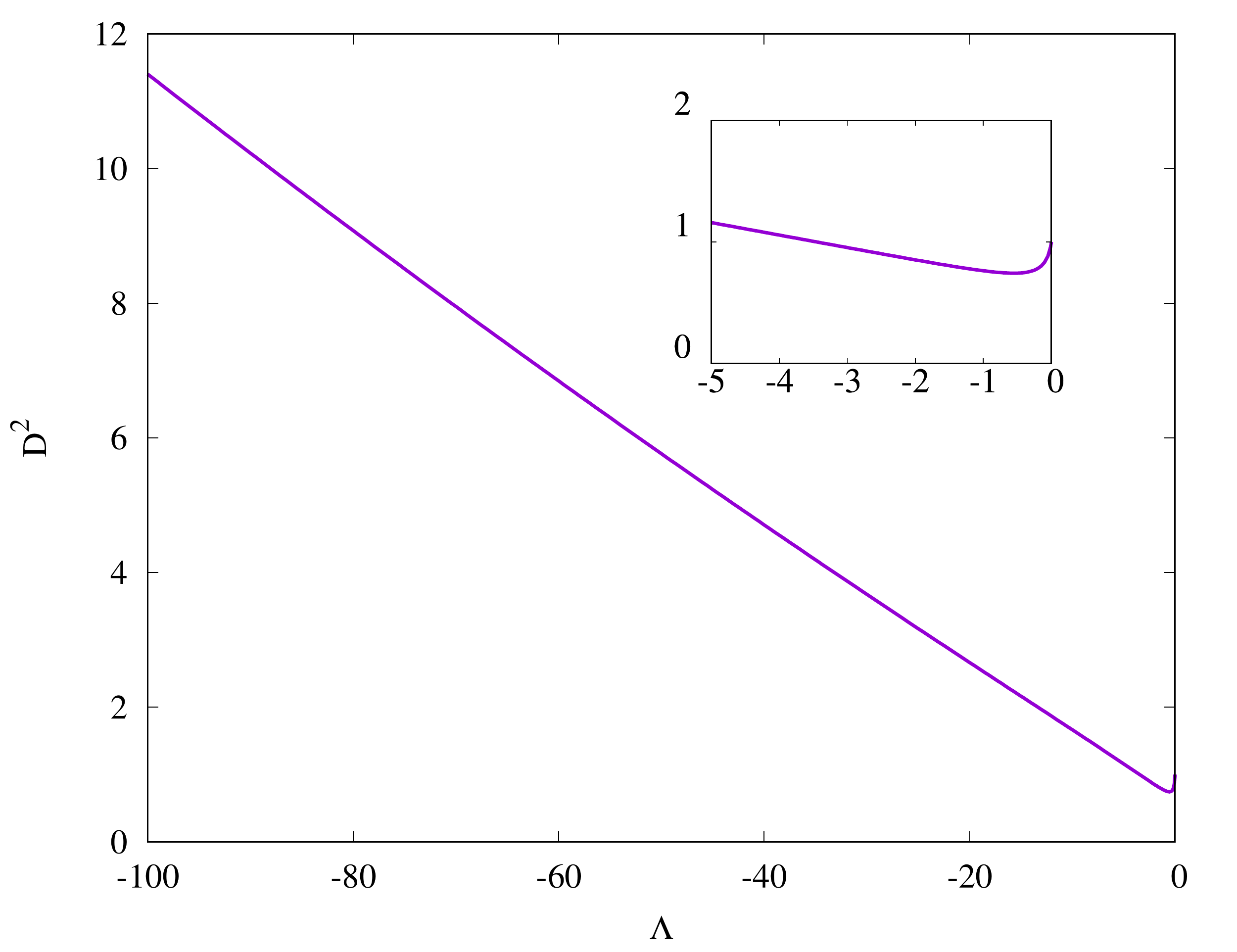}
(b) \includegraphics[angle =-90,scale=0.3]{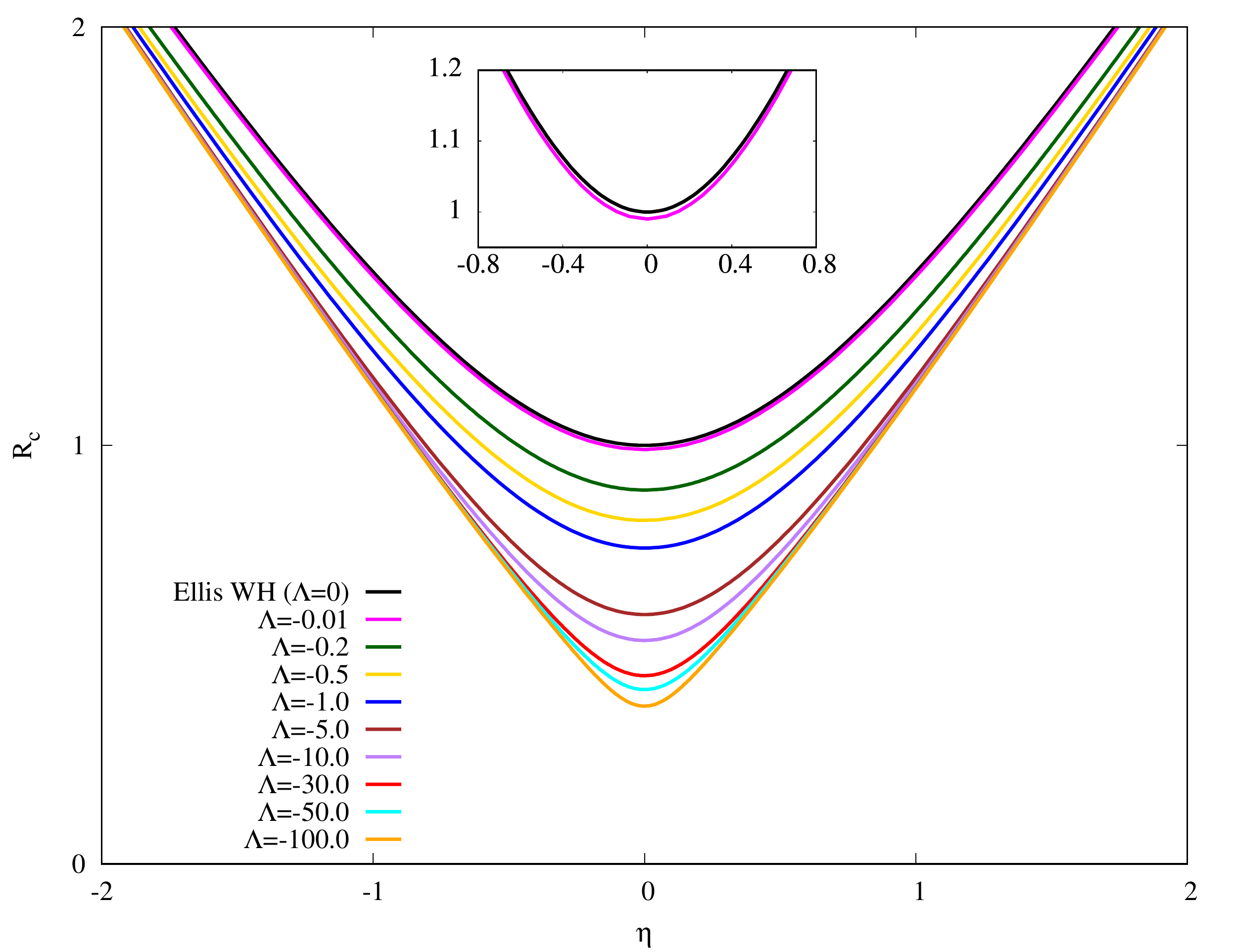}
}
\mbox{
(c)
\includegraphics[angle =-90,scale=0.3]{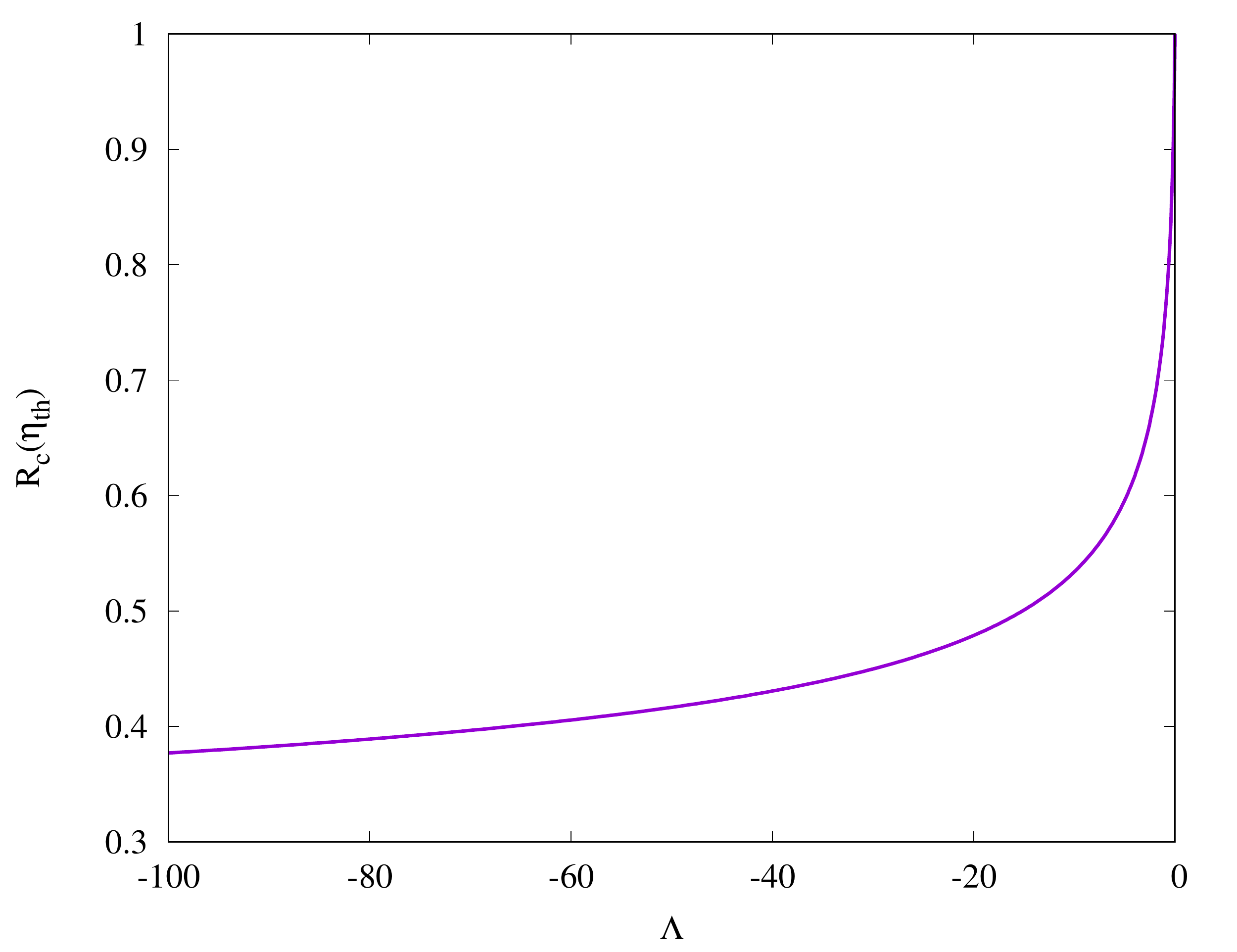}
(d)
\includegraphics[angle =-90,scale=0.3]{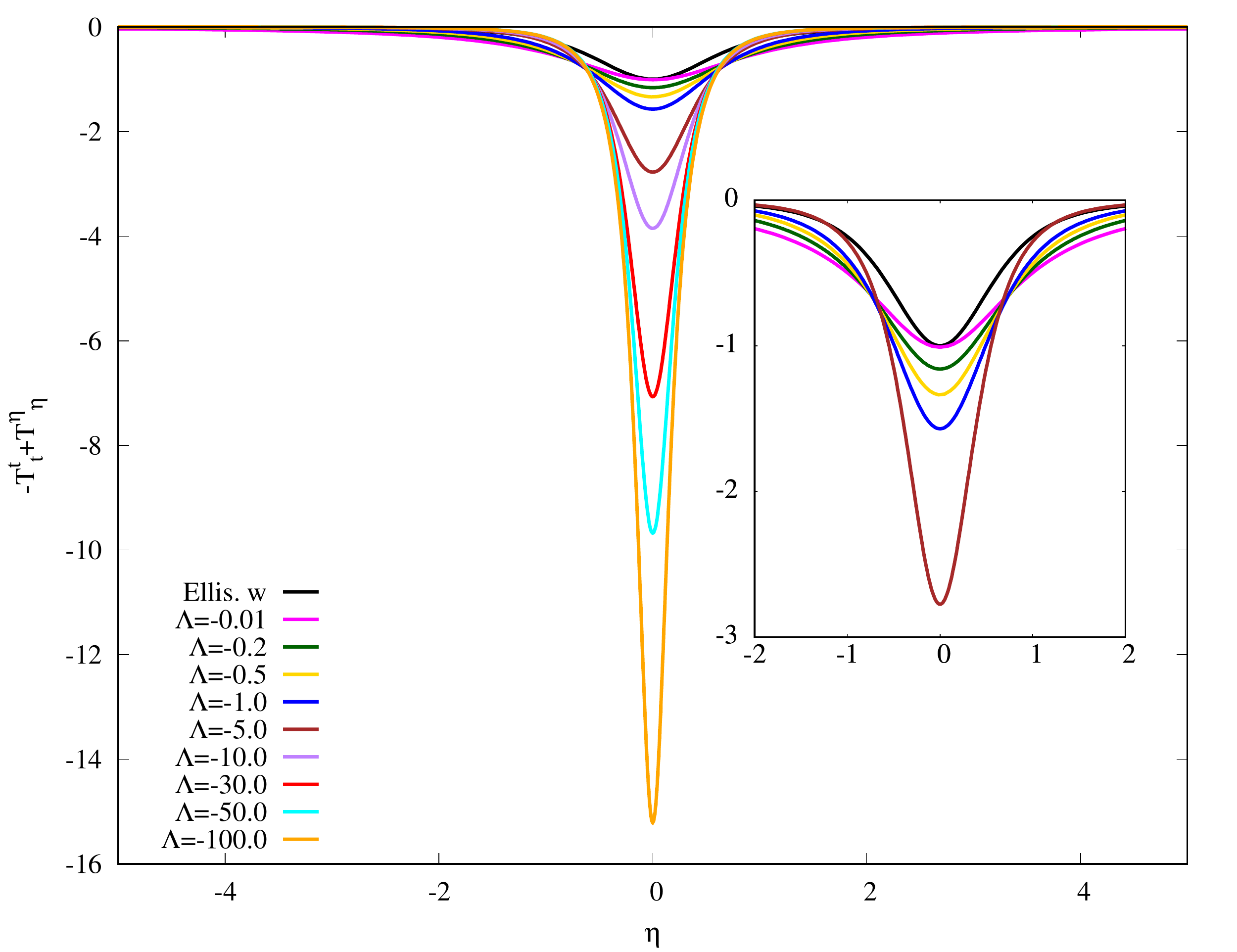} 
}
\caption{Properties of wormhole solutions: (a) phantom field constant $D^2$ vs $\Lambda$; (b) circumferential radius $R_c$ vs radial coordinate $\eta$ for several values of $\Lambda$; (c) circumferential throat radius vs $\Lambda$;  (d) NEC violation vs radial coordinate $\eta$ for several values of $\Lambda$.}
\label{plot_property}
\end{figure}

\begin{figure}[t!]
\centering
\mbox{ 
(a)  \includegraphics[angle =-90,scale=0.3]{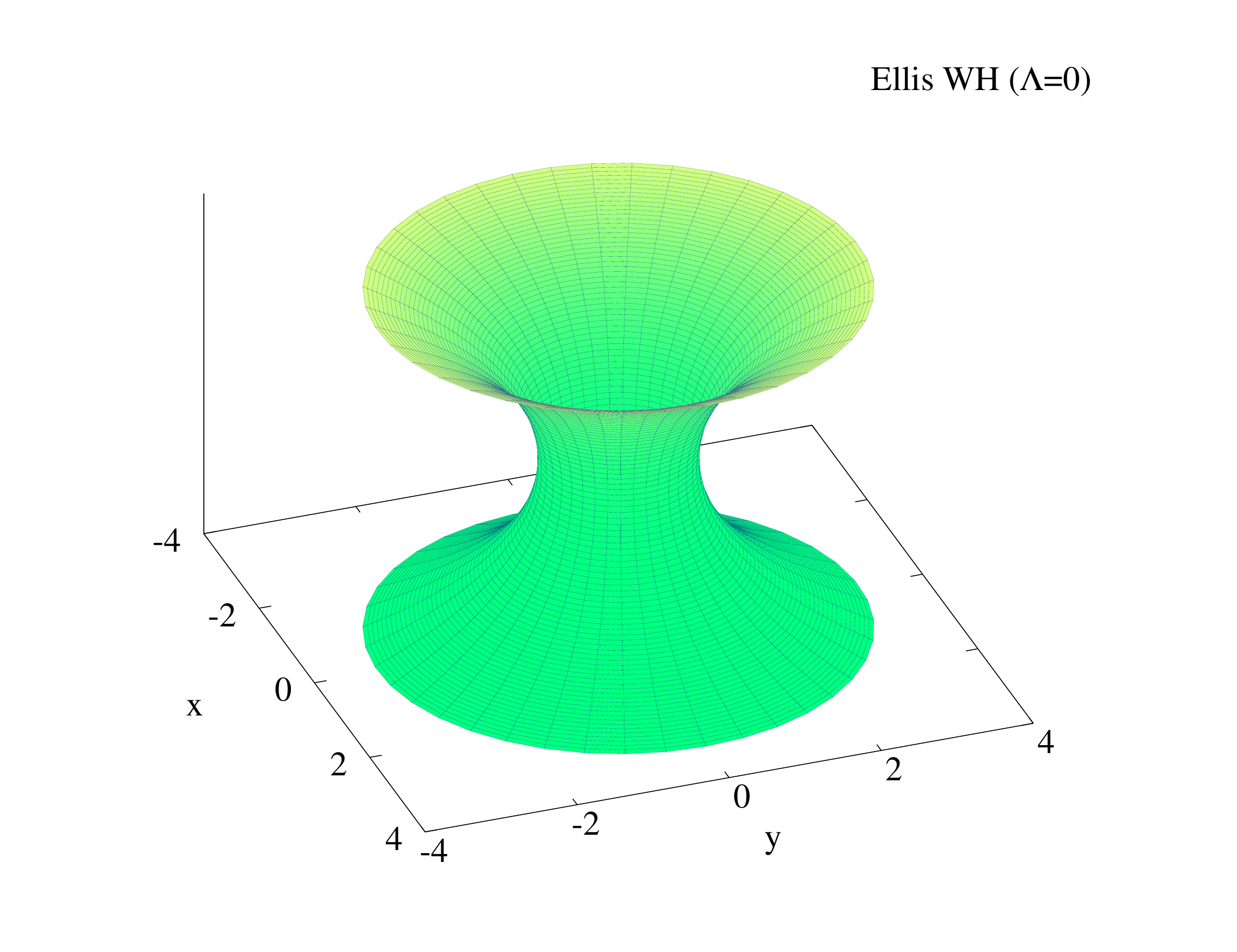}
(b) \includegraphics[angle =-90,scale=0.3]{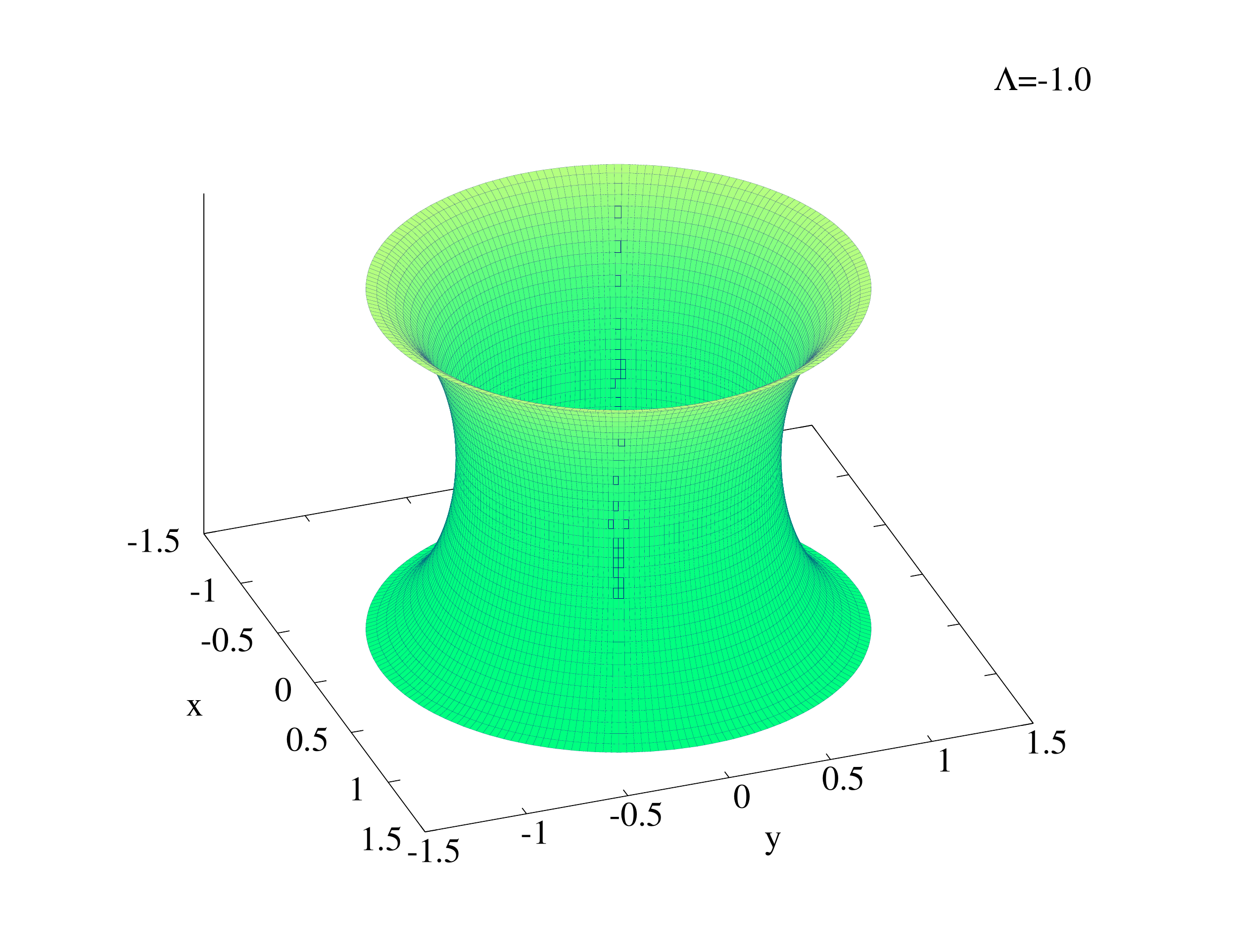}
}
\mbox{
(c)
\includegraphics[angle =-90,scale=0.3]{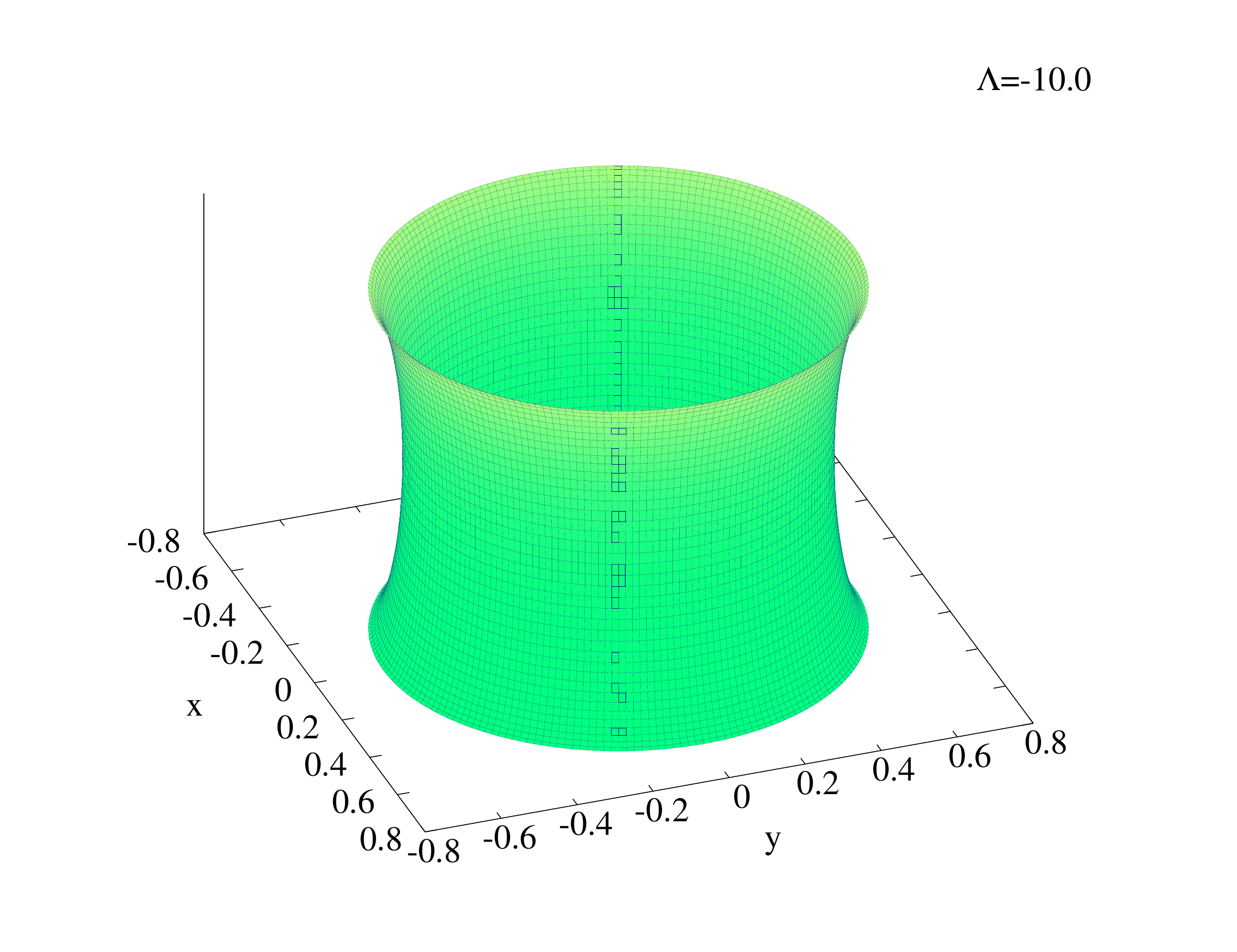}
(d)
\includegraphics[angle =-90,scale=0.28]{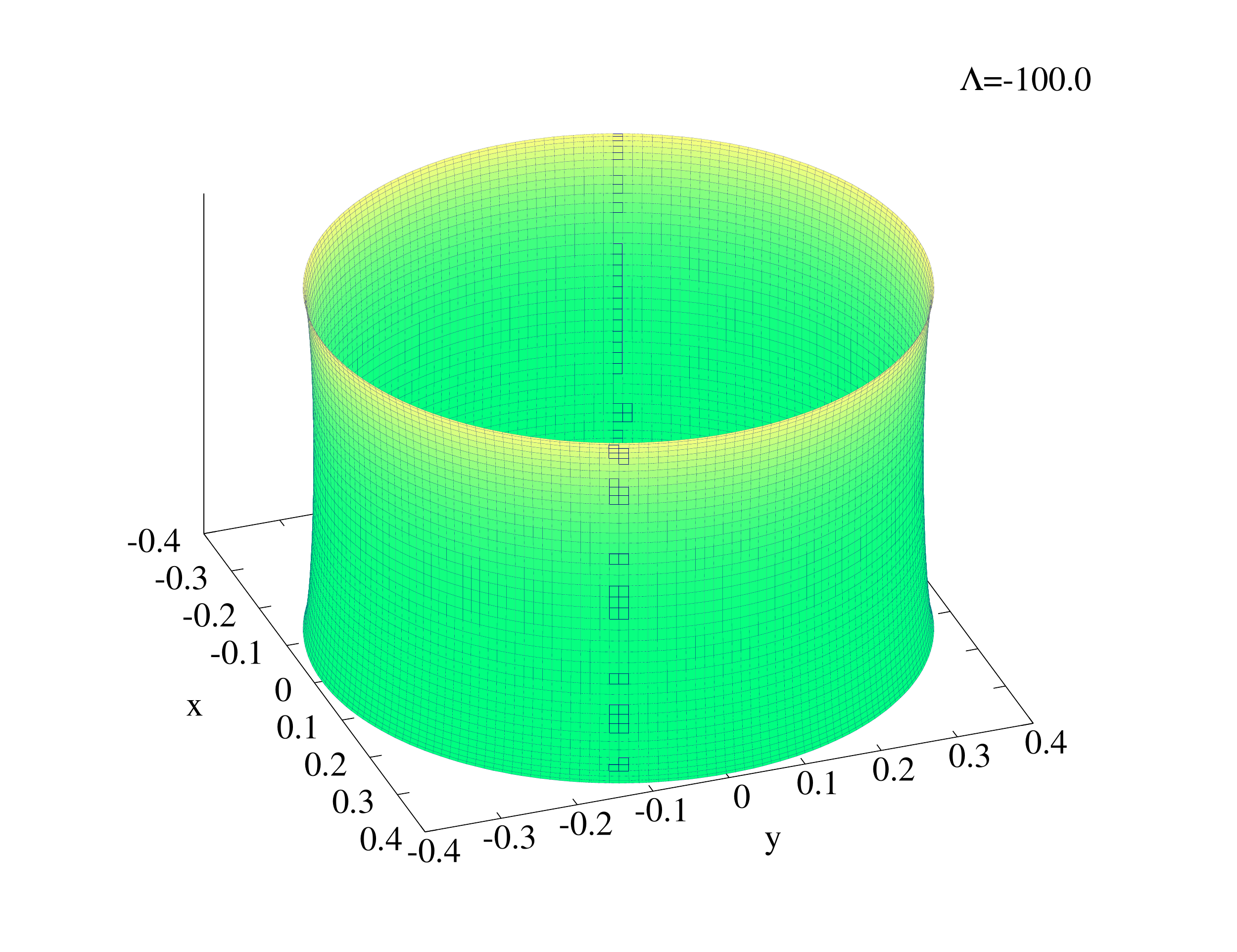} 
}
\caption{Isometric embeddings of wormhole solutions for several values of $\Lambda$: (a) $\Lambda=0.0$; 
(b) $\Lambda=-1.0$; (c) $\Lambda=-10.0$; (d) $\Lambda=-100.0$.}
\label{plot_3d}
\end{figure}

We have systematically constructed asymptotically AdS wormhole solutions for values of the cosmological
constant $\Lambda$ in the range $-100 \leq \Lambda \leq 0$. These wormholes are symmetric
with respect to $\eta \rightarrow -\eta$. In the limit $\Lambda \to 0$  the asymptotically flat Ellis wormhole
is obtained, where $F=p=1$.
We illustrate our results in Figs.~\ref{plot_metric}, \ref{plot_property} and \ref{plot_3d}.

The metric components $g_{tt}$ and $g_{\eta \eta}$ are shown in Figs.~\ref{plot_metric}(a) and (b), respectively,
for a set of values of the cosmological constant, including the asymptotically flat case (black).
As $\Lambda$ decreases from zero, the asymptotic behavior changes to AdS. This is reflected in the asymptotic
$\sim \eta^2$ and $\sim \eta^{-2}$ dependence of $g_{tt}$ and $g_{\eta \eta}$, respectively.
The component $g_{tt}$ has its global maximum at $\eta=0$. As $\Lambda$ decreases from zero, 
the global maximum of $g_{tt}$ increases slightly from $-1$ to a maximal value, and then decreases with
further decreasing $\Lambda$, as demonstrated in Fig.~\ref{plot_metric}(c).

The metric component $g_{\eta \eta}$ is shown in Fig.~\ref{plot_metric}(b). When $\Lambda$ decreases from zero it develops a local minimum at $\eta=0$. This local minimum is, however, surrounded symmetrically by two degenerate maxima. As $\Lambda$ decreases further the maxima move toward $\eta=0$, until they merge with the minimum. Finally a single maximum at $\eta=0$ remains, which decreases in size with further decreasing $\Lambda$.

In Fig.~\ref{plot_property} we address some properties of the asymptotically AdS wormhole solutions.
In Fig.~\ref{plot_property}(a) we exhibit the phantom field constant $D^2$ versus the cosmological constant $\Lambda$. When $\Lambda=0$, $D$ represents the phantom field charge, which is unity (given the scaling relations (\ref{scaling})) for a massless Ellis wormhole. With decreasing $\Lambda$, the constant $D^2$ at first decreases slightly to a minimum value, and then increases almost linearly as $\Lambda$ decreases further. 

Turning to the geometric properties of the wormholes, we show the circumferential radius $R_c$ versus the radial coordinate $\eta$ in Fig.~\ref{plot_property}(b) for several values of $\Lambda$. $R_c$ has a single minimum at $\eta=0$, from where it rises monotinically toward infinity on both sides. This minimum corresponds to the throat of the respective wormhole. Thus the wormholes possess a single throat. 

The circumferential radius $R_c$ of the throat is shown versus the cosmological constant in Fig.~\ref{plot_property}(c). The throat radius is largest, when $\Lambda=0$, which corresponds to the Ellis wormhole. The throat radius decreases monotonically as $\Lambda$ decreases. Note that the circumferential coordinate $R_c$ tends to the modulus of the radial coordinate $\eta$. 

The violation of the NEC as expressed via condition (\ref{necvio}) is demonstrated in Fig.~\ref{plot_property}(d). The violation is minimal when $\Lambda=0$, and thus for the Ellis wormhole. The NEC violation increases significantly at the throat when $\Lambda$ decreases. 

The wormhole throat can be visualized clearly with the help of embedding diagrams as shown in Fig.~\ref{plot_3d}. The figures also demonstrate that the size of the throat radius decreases as $\Lambda$ decreases. (Note the change of the grid size.)  


\section{\label{sec:lin}Linear Stability}

The investigation of the stability of wormholes is of considerable relevance. It is well-known that the asymptotically flat static Ellis wormholes in GR possess an unstable radial mode \cite{Gonzalez:2008wd,Gonzalez:2008xk,Torii:2013xba}. Therefore, we will now address the stability of static Ellis wormholes with AdS asymptotics by studying radial perturbations on these new background solutions. 

We start by introducing the following Ansatz for the line element
\begin{equation}
 ds^2 = - e^{ \nu(t,\eta) } N(\eta) dt^2 + e^{ -\lambda(t,\eta) } \frac{ p(\eta)}{ N(\eta) }  d\eta^2 + p(r) h(\eta) e^{ -\sigma (t,\eta)}  (d \theta^2+\sin^2 \theta d\varphi^2) \,,
\end{equation}
where
\begin{align}
 \nu (t,\eta) =  \nu_0 (\eta) + \epsilon \nu_1 (\eta) e^{-i \omega t}    \,, &   \quad \lambda (t,\eta) =  \lambda_0 (\eta) + \epsilon \lambda_1 (\eta) e^{-i \omega t}  \,, \quad  \sigma (t,\eta) =  \sigma_0 (\eta) + \epsilon \sigma_1 (\eta) e^{-i \omega t}  \,,
 \end{align}
and the relation with the background metric is
\begin{equation}
 e^{\nu_0} = e^{-\lambda_0} = e^{-\sigma_0}= F \,.
\end{equation}
For the phantom field we employ the Ansatz
 \begin{align}
 \Psi(t,\eta) &= \psi(\eta) + \epsilon \Psi_1 (\eta)  e^{-i \omega t}\,.
\end{align}
A mode with eigenvalue $\omega^2$ is unstable and increases exponentially when $\omega^2 < 0$.

Inserting the Ans\"atze into the scalar field equation,
we find 
\begin{equation} \label{KGeqn_perturb}
  \Psi_1''+ \left( \frac{p'}{2 p} + \frac{2 \eta}{h} - \frac{2 \Lambda \eta}{3 N}  \right) \Psi_1' + \omega^2 \frac{p \Psi_1 }{F^2 N^2} = 0 \,,
\end{equation}
when choosing the simple gauge--fixing 
\begin{equation}
 \lambda_1  =  -\nu_1 + 2 \sigma_1 \,.
 \label{gauge}
\end{equation}
Inserting the Ans\"atze into the Einstein equations, we obtain the following set of first--order ODEs when making use again of the gauge--fixing (\ref{gauge})
\begin{align}
 \nu_1' &= \frac{q_1}{6 p h N q_0} \nu_1 - \frac{q_2}{18 p h F N^2 q_0} \sigma_1  -   \frac{2 D F (-6 \eta p N- 3 h N p'+2 \Lambda \eta p h)}{ 3 h \sqrt{p} N^2 q_0}    \Psi_1  +   \frac{4 D F \sqrt{p} }{  N q_0 } \Psi_1'  \,, \label{odeh0r}  \\
   \sigma_1 '&= \left( -\frac{p'}{2 p}  - \frac{\eta}{h}  + \frac{F'}{2 F} \right) \nu_1  + \left( \frac{\eta}{h} - \frac{\Lambda \eta}{3 N} + \frac{p'}{2 p}  \right) \sigma_1  -\frac{2 D}{h N \sqrt{p}} \Psi_1  \,,  \label{odeh2r}
\end{align}
where
\begin{align} 
 q_0 &=- 2 \eta p F - h F p' + p h F'  \,, \\
 q_1 &= -12 h F p^2 + 12 \eta^2 p^2 F N + 3 F N h^2 p'^2 + 12 \Lambda h^2 p^3  - 4 \Lambda \eta^2 h F p^2 + 12 \eta p N h F p'  - 3 p N h^2 F' p' - 6 \eta h N p^2 F'     \nonumber \\ 
& \quad + 2 \Lambda \eta  p^2 h^2 F' - 2 \Lambda \eta p F h^2 p'     \\
 q_2 &=  36 \eta^2 p^2 F^2 N^2 + 9 N^2 h^2 F^2 p'^2  - 36 h N p^2 F^2 -12 \Lambda \eta p N h^2 F^2 p'  + 4 \Lambda^2 \eta^2 p^2 h^2 F^2   + 36 p^3 h^2 \omega^2 + 36 \eta h p F^2 N^2 p'   \nonumber \\ 
& \quad  -24 \Lambda \eta^2 h N p^2 F^2 + 72 \Lambda F N h^2 p^3  \,.
\end{align}

We now first analyze Eq.~\eqref{KGeqn_perturb} for the scalar field perturbation $\Psi_1$. By multipling Eq.~\eqref{KGeqn_perturb} by a factor $h N \sqrt{p}$ and making use of the product rule, we can combine the two derivative terms on the left hand side to obtain
\begin{equation}
 \left(  h N \sqrt{p} \Psi_1'  \right)' + \omega^2 \frac{h p^{3/2}  \Psi_1 }{N F^2 }  = 0 \,.
\end{equation}
We then multiply the new ODE by $\Psi_1$, and integrate by parts over the full interval $(-\infty,\infty)$. Hence we obtain
\begin{equation}
 \left(  h N \sqrt{p} \Psi_1 \Psi_1' \right) \Big|_{-\infty}^{\infty} = \int_{-\infty}^{\infty}  h \sqrt{p} \left(  N  \Psi_1'^2    - \omega^2 \frac{ p  \Psi_1^2 }{N F^2 }  \right)  d \eta  \,.
\end{equation}
Since we require $\Psi$ to be normalizable, the left hand side of this equation vanishes. Therefore the integral on the right hand side must also vanish. However, the integrand is non-negative if an unstable mode $(\omega^2 < 0)$ exists. Consequently $\Psi_1$ has to be zero, so that the integral will vanish identically. 

Next we consider Eqs.~\eqref{odeh0r} and \eqref{odeh2r}. They can be combined into a single master equation which is Schr\"odinger--like when written in terms of the perturbation function $Z=G_s \sigma_1$,
\begin{equation} \label{master_eqn}
 \frac{d^2 Z}{d r_{*}^2} + (\omega^2 - V_R(\eta) ) Z = 0 \,, 
\end{equation}
where $r_{*}$ is the tortoise coordinate and $V_R(\eta)$ is the radial effective potential,
\begin{equation}
 \frac{d r_{*}}{d \eta} = \frac{\sqrt{p}}{F N} \,, \quad V_R(\eta) = \frac{Q_1}{12 h^2 p^3 Q_0} \,,   
\end{equation}
with
\begin{align}
 \frac{1}{G_s}\frac{dG_s}{d\eta} &= -\frac{1}{2 \eta^2 h p N F q_0} \times \left( -4 \eta^3 h p N F^2 p' + 12 F N h^2 p^3 -12 F h^2 p^3 - 4 \eta^4 N p^2 F^2+4 \eta^2 h p^2 F^2 - N \eta^2 h^2 F^2 p'^2  \right.  \nonumber  \\
\qquad \qquad &   \qquad \left. - \eta^2 N h^2 p^2 F'^2 + 2 \eta^2 p F N h^2 p' F' + 4 \eta^3 h F N p^2 F' \right) \,, \\ 
 Q_0 &= p^2 h^2 F'^2 - 4 \eta h F p^2 F' - 2 p F h^2 p' F' + 4 \eta^2 p^2 F^2 + 4 \eta p h F^2 p' + h^2 F^2 p'^2 \,, \\ 
Q_1 &= -48 N^2 \eta^4 p^4 F^4 - 3 N^2 h^4 F^4 p'^4  + 96 \Lambda^2 F^2 h^4 p^6 - 192 \Lambda h^3 F^3 p^5  + 144 \eta^2 h^2 N^2 p^3 F^3 F' p' +   72 \eta p^2 N^2 h^3 F^3 F' p'^2  \nonumber  \\
& \quad + 48 \eta \Lambda N F^3 h^3 p^4 p' - 72 \eta F^2 N^2 h^3 p^3 p' F'^2  - 3 N^2 p^4 h^4 F'^4 + 36 N F^2 h^3 p^4 F'^2  - 64 \Lambda^2 \eta^2 h^3 F^3 p^5  \nonumber \\
& \quad + 64 \Lambda \eta^2 h^2 p^4 F^4 - 12 N p^2 h^3 F^4 p'^2 - 48 h N \eta^2 p^4 F^4 + 12 p N^2 F^3 h^4 F' p'^3   - 48 \eta N h^2 p^3 F^4 p'    \nonumber \\    
& \quad -48 \eta N h^2 F^3 p^4 F' -  24 N F^3 h^3 p^3 p' F' + 32 \eta \Lambda^2 F^2 h^4 p^5 F' - 36 \Lambda F N h^4 p^5 F'^2  -32 \eta \Lambda^2 F^3 p^4 h^4 p'  \nonumber \\
& \quad  - 32 \Lambda \eta F^3 h^3 p^4 F'  + 32 \Lambda \eta h^3 p^3 F^4 p' +12 F N^2 p^3 h^4 p' F'^3 + 24 \eta F N^2 h^3 p^4 F'^3  - 72 \eta^2 h^2 F^2 N^2 p^4 F'^2  \nonumber \\
& \quad  + 48 \Lambda \eta^2 N h^2 F^3 p^5 + 96 h N^2 \eta^3 F^3 p^4 F'  -96 h N^2 \eta^3 p^3 F^4 p' -72 \eta^2 h^2 N^2 p^2 F^4 p'^2  + 12 \Lambda N h^4 F^3 p^3 p'^2  \nonumber \\
& \quad- 18 p^2 N^2 F^2 h^4 p'^2 F'^2-
24 \eta p N^2 h^3 F^4 p'^3  + 48 \Lambda \eta N F^2 h^3 p^5 F'  + 24 \Lambda N F^2 h^4 p^4 F' p' + 96 h^2 p^4 F^4    \,, 
\end{align}
where we have used $F''$ and $p''$ to simplify the above expressions.

\begin{figure}[t!]
\centering
\mbox{
(a)
 \includegraphics[angle =-90,scale=0.3]{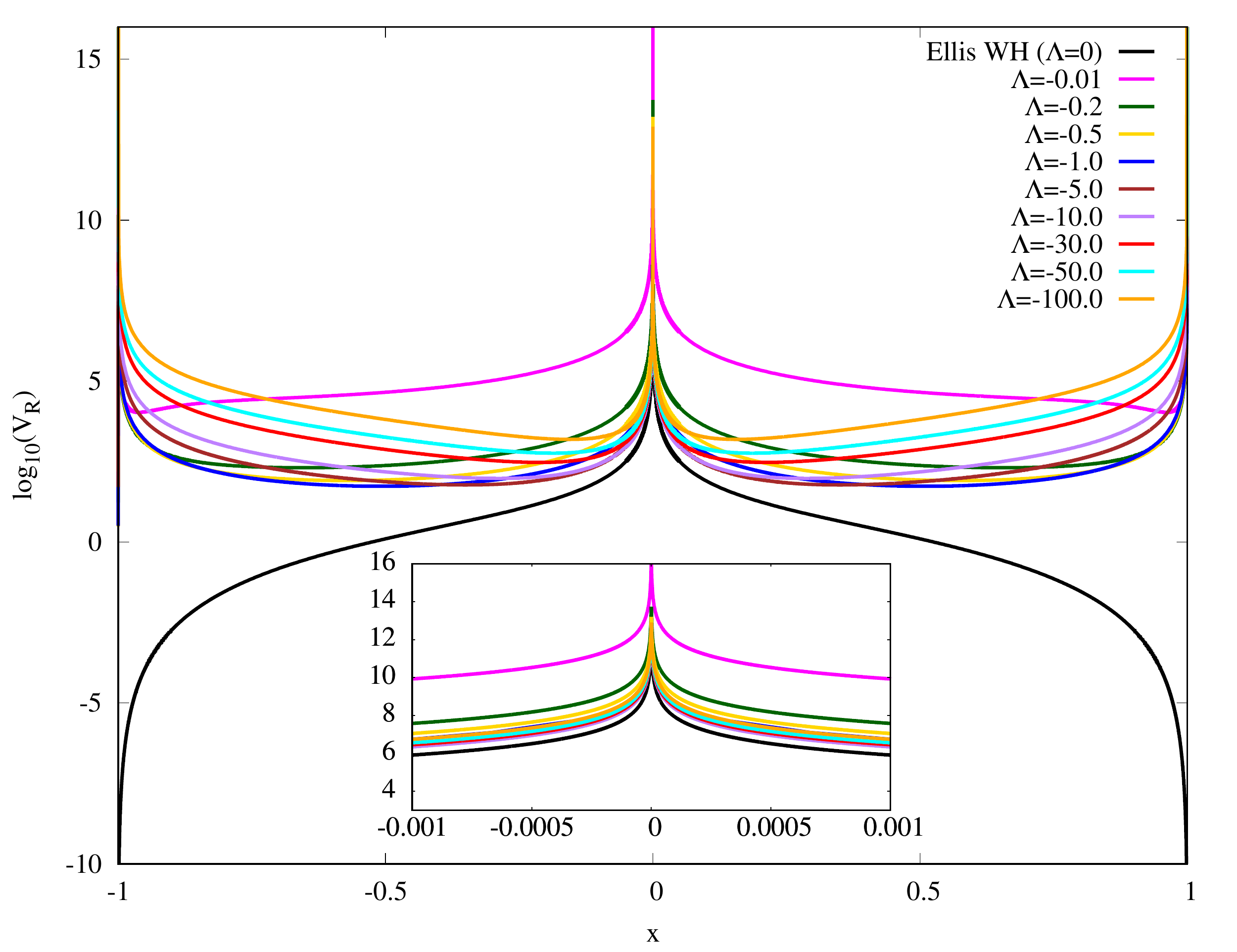}
(b)
 \includegraphics[angle =-90,scale=0.3]{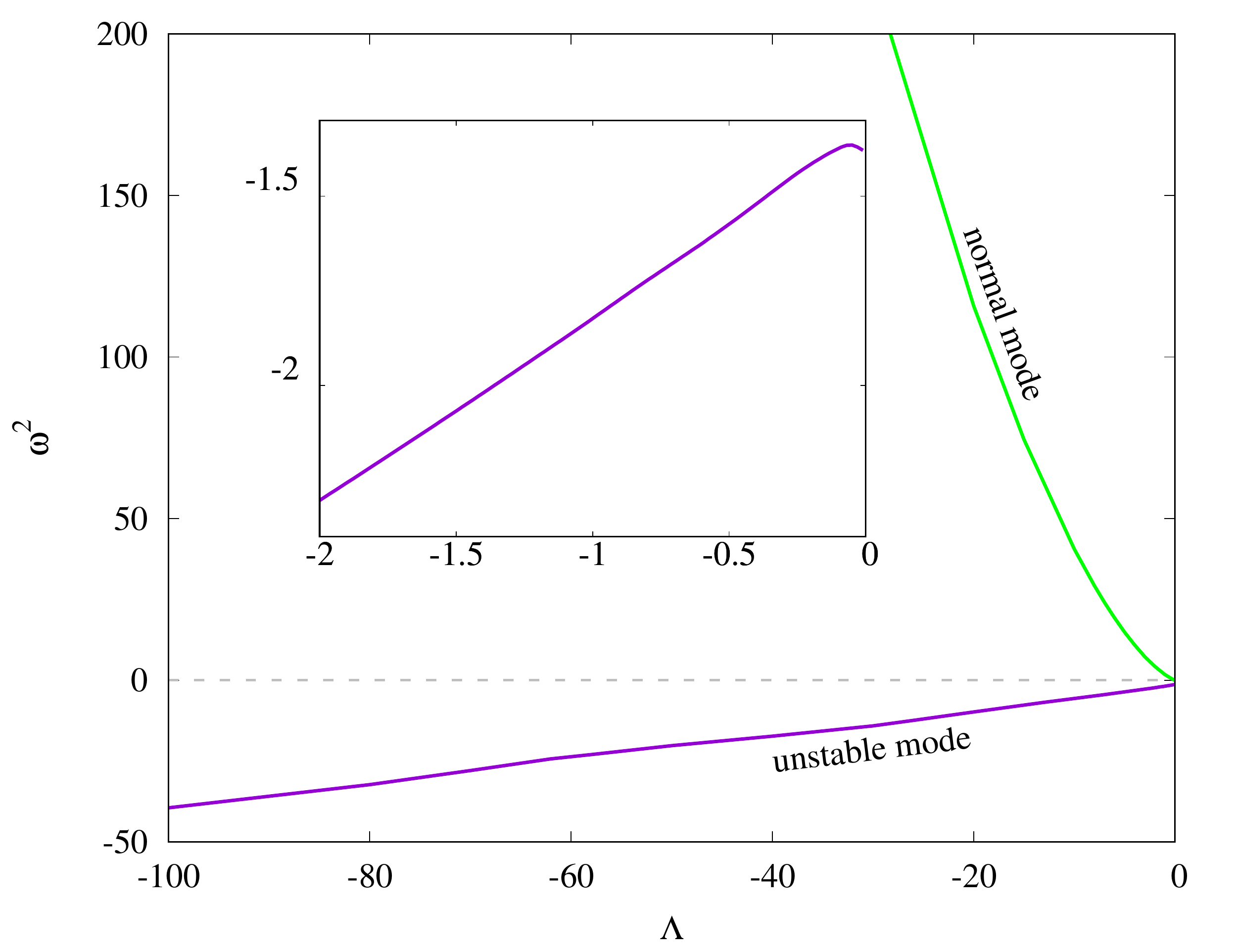}
 }
\caption{(a) Radial effective potential $V_R$ vs compactified radial coordinate $x$ for several values of $\Lambda$; (b) eigenvalue $\omega^2$ of the unstable mode vs $\Lambda$.} 
\label{plot_w2}
\end{figure}

The radial effective potential is illustrated in Fig.~\ref{plot_w2}(a) versus the compactified radial coordinate $x$ for several values of the cosmological constant. It is symmetric with respect to $\eta \to -\eta$, exhibiting {symmetric minima. The larger $\Lambda$ is, the farther away from the throat are these minima of the potential. 
	
Similarly to what happens in the asymptotically flat clase, the potential diverges at the throat $\eta=0$, as indicated in Fig.~\ref{plot_w2}(a). The analytical expansion of the potential around the throat shows that it diverges like $\eta^{-2}$, 
\begin{equation}
 V_R (\eta) = \frac{2 F_0^2}{p_0 \eta^2} + O(\eta^0) \,.
\end{equation}

However, a difference with respect to the asymptotically flat case is that the potential also diverges at the two spatial infinities when the cosmological constant is not zero, as also indicated in Fig.~\ref{plot_w2}(a). The expansion of the potential at infinity shows that the potential diverges like $\eta^2$, 
\begin{equation}
 V_R (\eta) = \frac{2}{9} F_\infty \Lambda^2 \eta^2 + O(\eta) \,.
\end{equation}
}

We compute the unstable mode numerically by using COLSYS to solve Eq.~\eqref{master_eqn}, which is an eigenvalue problem with $\omega^2$ as the eigenvalue. In general the perturbation function $Z$ does not vanish at the infinities (this is the case because these are spherical perturbations of an asymptotically AdS configuration). However the perturbation equation requires that the derivative of the perturbation function vanishes at the boundaries, $\partial_{\eta}Z(-\infty)=\partial_{\eta}Z(\infty)=0$. We impose these as boundary conditions. Since Eq.~\eqref{master_eqn} is homogeneous, in order to obtain a nontrivial and normalizable solution for $Z$, we introduce an auxiliary equation $\frac{d}{d\eta} \omega^2=0$, that allows us to impose the condition $Z(\eta_p)=1$ at some point $\eta_p$ which we typically choose to be above the throat. The eigenvalue $\omega^2$ is found when $Z$ satisfies all the asymptotic boundary conditions. With this method, the estimated error of the modes reported in the following is typically $1\%$ or smaller.

We exhibit the eigenvalue $\omega^2$ versus the cosmological constant in Fig.~\ref{plot_w2}(b). We show two different modes. In purple we show the unstable mode with $\omega^2<0$. This unstable mode is found for any value of the cosmological constant, and in particular, as $\Lambda \to 0$, it tends to the value of the unstable mode for the asymptotically flat mass-less Ellis solution (as seen in the inset). Hence we conclude that the change of asymptotics does not stabilize the wormhole, since the unstable mode of the asymptotically flat solution can be continued smoothly for arbitrary values of cosmological constant. In fact, note that the larger the magnitude of the cosmological constant (i.e., the shorter the AdS length) the larger the value of the unstable mode becomes.

In this figure we also show in green a nodeless normal mode with $\omega^2>0$. This mode corresponds to stable perturbations. Because of the AdS asymptotics, gravitational waves are reflected by the conformal boundary, acting in practice like a box. This is different to the asympotically flat case, where the perturbations oscillate while being damped exponentially in time (quasinormal modes). With AdS asymptotics however, the oscillation is not damped, but reflected at the boundaries, and it is possible to find normal modes like the one we include in the figure. Note that the frequency grows as we decrease the AdS length. And in the limit $\Lambda\to 0$, it vanishes with $\omega^2=0$.   

\section{\label{sec:con}Conclusion and Outlook}

We have constructed numerically the globally regular solutions of static Ellis wormholes with AdS asymptotics. The wormhole solutions are symmetric with respect to $\eta \to -\eta$ and consequently massless. Our numerics indicate that these solutions exist for any value of the negative cosmological constant, although in this work we have focused the discussion on the range $-100 \le \Lambda <0$. The AdS asymptotics of the metric is reflected in the corresponding dependence of the metric components, where the $g_{tt}$ component behaves as $\sim \eta^2$ at spatial infinity, while $g_{\eta \eta}$ behaves as $\sim -\eta^{-2}$.
 
The wormholes possess a single throat which, because of the symmetry and the radial coordinate that we have used, is located at the position $\eta=0$. By decreasing $\Lambda$, the (scaled) circumferential radius of the throat decreases monotonically, approaching zero in the limit of infinite coupling constant. In this case the circumferential coordinate tends towards the modulus of the radial coordinate $\eta$. The wormholes violate the null energy condition, as implied by the presence of a phantom field.

In addition, we have also studied stability of these wormholes against spherically symmetric perturbations. Using the same approach as for the asymptotically flat Ellis wormholes, we have shown that the asymptotically AdS wormholes are also unstable against radial linear perturbations. The unstable mode tends to the value of the asymptotically flat case when $\Lambda\to0$, and it increases in magnitude as the AdS length is decreased. We have also shown the existence of normal modes, which are allowed by the box-like properties of the spacetime.

Since asymptotically flat Ellis wormholes can rotate \cite{Kashargin:2007mm,Kashargin:2008pk,Kleihaus:2014dla,Chew:2016epf,Dzhunushaliev:2013jja}, it will be interesting to construct the rotating generalizations of these new Ellis wormholes with AdS asymptotics. 

Furthemore, if a wormhole solution is allowed by a string-inspired model in the asymptotically AdS background, it will open a new window toward the information loss paradox. In this model, can the philosophy of the ER=EPR conjecture still be true? What is the physical meaning of the violation of the causality in such a theory? Can the same method of \cite{Almheiri:2019hni} be applicable in this theory? We leave these interesting questions for future projects.

\section*{Acknowledgement}

XYC and DY were supported by the National Research Foundation of Korea (Grant No.: 2018R1D1A1B07049126).
JLBS and JK gratefully acknowledge support by the
DFG Research Training Group 1620 {\sl Models of Gravity}
and the COST Action CA16104. JLBS would like to acknowledge support from the DFG project BL 1553, and the FCT projects PTDC/FISOUT/28407/2017 and PTDC/FIS-AST/3041/2020.

\end{document}